\documentclass[journal]{IEEEtran}

\usepackage[normalem]{ulem}
\usepackage{subfigure}
\usepackage{amsmath}
\usepackage{graphicx}
\usepackage{subfigure}
\usepackage{amssymb}
\usepackage{array}
\usepackage{accents}
\usepackage{bbm}
\usepackage{mathtools}

\begin{document}
\title{Control and Communication Protocols that Enable Smart Building Microgrids}
%
%
%

\author{Bowen~Zhang,~\IEEEmembership{Student Member,~IEEE,}
        John~Baillieul,~\IEEEmembership{Fellow,~IEEE}
\thanks{Manuscript received ********; revised ********. Research reported here was supported by NSF grant 1038230.}
\thanks{\textbf{B. Zhang} is with the Division of Systems Engineering, Boston University, Boston,
MA, 02215 USA e-mail: bowenz@bu.edu.}
\thanks{\textbf{J. Baillieul} is with the Department of Mechanical Engineering, Department of Electrical and Computer Engineering, and 
Division of Systems Engineering, Boston University, Boston,
MA, 02215 USA e-mail: johnb@bu.edu.}}

\IEEEspecialpapernotice{-(Invited Paper)}

\maketitle

\begin{abstract}
Recent communication, computation, and technology advances coupled with climate change concerns have transformed the near future prospects of electricity transmission, and, more notably, distribution systems and microgrids. Distributed resources (wind and solar generation, combined heat and power) and flexible loads (storage, computing, EV, HVAC) make it imperative to increase investment and improve operational efficiency. Commercial and residential buildings, being the largest energy consumption group among flexible loads in microgrids, have the largest potential and flexibility to provide demand side management. Recent advances in networked systems and the anticipated breakthroughs of the Internet of Things will enable significant advances in demand response capabilities of intelligent load network of power-consuming devices such as HVAC components, water heaters, and buildings. In this paper, a new operating framework, called packetized direct load control (PDLC), is proposed based on the notion of quantization of energy demand. This control protocol is built on top of two communication protocols that carry either complete or binary information regarding the operation status of the appliances. We discuss the optimal demand side operation for both protocols and analytically derive the performance differences between the protocols. We propose an optimal reservation strategy for traditional and renewable energy for the PDLC in both day-ahead and real time markets. In the end we discuss the fundamental trade-off between achieving controllability and endowing flexibility.

\end{abstract}


%
\IEEEpeerreviewmaketitle

\section{Introduction: Energy Evolution}
The traditional power system is under significant evolution to the \textit{smart grid} that is driven by technology, economics, as well as policy needs \cite{varaiya2011smart}. Technology advances provide the foundation that enables the full realization of the smart grid. One such advancement is the prevailing installation of distributed generation (DG) and renewable energy resources all around the world. The European Union is the leader in developing and deploying renewable generation to reduce the dependence on imported energy and greenhouse gas emissions. The EU has set a goal to reach 20\% share of renewable energies in gross energy consumption by 2020. In United States, 38 states have long term renewable portfolio standards and 14 states have installed more than 1,000 MW of wind power \cite{AWEAGOV}. It is expected that DG and renewable deployment will have an annual growth of 2.5\% per year until 2040 \cite{eia2013}. Parallel to the development of generation technologies is the profound structural changes in demand. Traditional passive demands are becoming intelligent and an increasing number of end users are participating in demand response programs that enhance grid reliability in presence of uncertain events such as failure of generators or transmission lines, renewable intermittency, etc. An extra level of reliability is guaranteed by the development of storage elements that are installed at the microgrid level to store and release energy in case of emergencies.

The information aspects of the power system, i.e. sensing, communication, and algorithmic approaches, are the key soft elements for improving grid intelligence. Wide-Area Measurement Systems (WAMS), also known as synchrophasors, are experiencing a tremendous era in the past decade. The Phasor Measurement Unit, known as the PMU, is broadly deployed in the transmission network to measure 60 Hz waveforms at fine resolution. Based on the concept of a phasor as introduced by Charles Proteus Steinmetz in 1893, \cite{steinmetz1893complex}, modern PMU's can provide synchronized data by using clock signals from GPS. Thus high-resolution measurements of real and reactive power flows taken at widely separated points in the grid can be compared in real time to provide high accuracy security assessment and disturbance localization \cite{chakrabortty2013introduction,6672612}. At the distribution and microgrid level, advanced metering infrastructure (AMI) integrates sensor, appliances, metering and data management systems to enable the information exchange between utilities and end users. These infrastructures are typically two-layer designed where the first layer is the wired or wireless communication established between meters and appliances, and the second layer is the Internet based communication between utilities and users.

Generation advancement, demand flexibility, and information technologies together are shaping the traditional power system, whose operation has been centralized for the past century. The success of the smart grid will depend heavily on the operation of distributed microgrids in which smart buildings are the most energy consuming parts, but at the same time having the highest potential and flexibility to manage. In this paper, our foremost focus is on the optimal operation of smart buildings assisted by various control and communication technologies. Section \ref{smart microgrid} presents a brief review of the development and operation of microgrids. Section \ref{smart building} introduces the most important flexible resource of the microgrid -- smart buildings. We present a brief literature review that addresses recent research on modelling, control, and energy efficiency aspects of the smart buildings. Inspired by the notion of packet switched communication system, we introduce a new operating concept, called the {\it energy packet}, to the existing direct load control (DLC) frameworks. The corresponding framework called {\it packetized direct load control} (PDLC) is discussed in Section \ref{energy packet and the pdlc}. The rest of the paper elaborates the performance opportunities of PDLC under two communication and control scenarios. The first scenario provides a baseline where the PDLC can smooth demand uncertainty and guarantee consumers' energy needs based on real time full state information. The second scenario considers constrained information passing when only binary state information is transmitted. We characterize the performance with these protocols in terms of both energy and monetary metrics in Section \ref{Quantifying Performance Degradation}. The impact of volatile resources to the system performance is discussed in Section \ref{Impact of Volatile Resources}. Section \ref{Optimal Hourly Energy Procurement} discusses the optimal energy portfolio procurement for both traditional and volatile resources. Section \ref{Control and Market Aspects for the Binary Information PDLC} addresses the control and market aspects of binary information PDLC. Conclusion and future directions are drawn in Section \ref{conclusion}.

\section{Smart Microgrids}\label{smart microgrid}
\begin{figure}[h]
\begin{center}
\includegraphics[width=8.4cm]{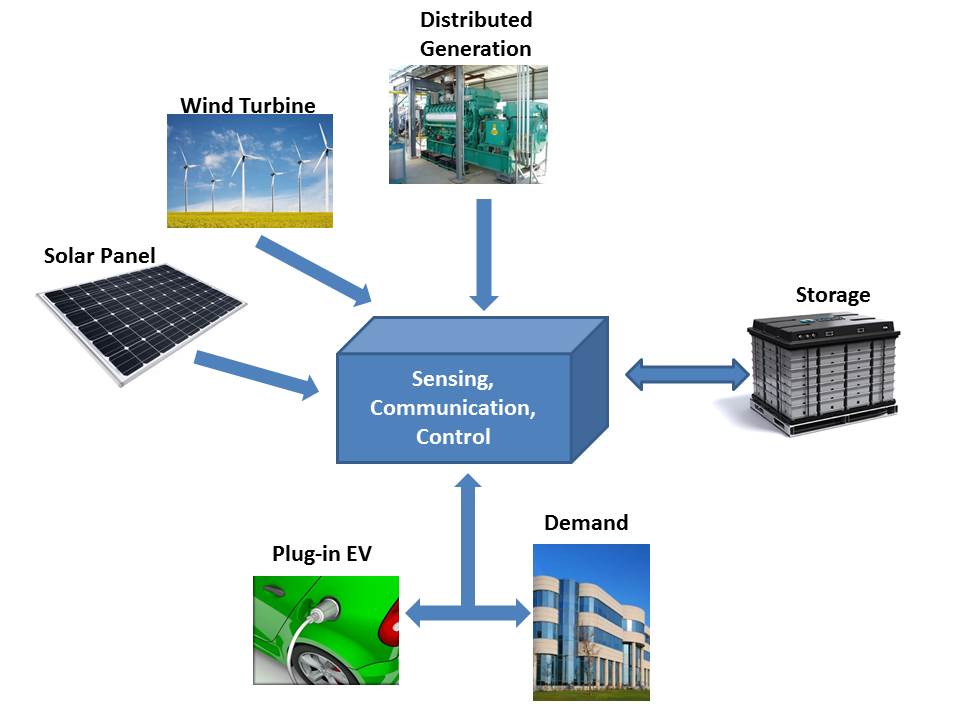}    
\caption{A typical microgrid is comprised of smart buildings, distributed and renewable generations, energy storage elements, and a central microgrid controller.} 
\label{fig: microgrid}
\end{center}
\end{figure}

The concept of microgrid is markedly different from te components of a the traditional power system composed of numerous generation units, transmission/distribution lines, and loads. Shown in Fig. \ref{fig: microgrid}, a microgrid is a localized distribution system composed of distributed generators, flexible loads, and energy storage elements that are all networked through advanced communication techniques and are controlled by the microgrid operator who is responsible for providing reliable and secure electricity service. Drivers of the growth of microgrids include regulation incentives, consumers' needs, and operation costs. 

The development of microgrids can facilitate the adoption of renewable resources \cite{jiayi2008review,1354758}. This is consistent with the regulatory goal of the US Government that an increasing proportion of renewable energy must be used to serve energy demand \cite{cal2020,mass2020}. For example, California requires that all sellers of electricity must serve their load with 33\% of electricity generated by wind by 2020. Massachusetts has committed to installed 2,000 MW of wind generation before 2020 as part of its plan to reduce greenhouse gas emissions. Microgrids can also enhance the resilience of the power system in order to withstand both physical and cyber attacks \cite{hahn2013cyber,kundur2011towards,hines2014smart}. When an major disturbance occurs, such as super-storms or floods, the localized microgrid can actively island itself from the main grid to operate using its distributed generators and energy storage elements. The detach of microgrid from the main system will also prevent the occurrence of large scale cascaded failure. In addition, microgrid operation is facilitated by a few advanced information technologies that are ready to use. The advent of Internet of Things allows the networked connectivity between the operator and the flexible loads such that communication signals are exchanged and control signals are executed \cite{galli2011grid,5622060,huang2010analytics}. Metering technologies together with data analytics enable the operator to gain better understanding of consumers preferences and behaviours. In addition, various software have been developed to maximize the economic use of resources and to minimize the costs of consuming electricity.

Parallel to the potential benefits that the microgrid will bring to power systems, research has been conducted to optimize the operation of microgrids such that maximum economic benefit can indeed be achieved. Renewable resources, whose generation outputs largely depend on the stochastic environments, need to be coordinated and controlled. One solution is to build hybrid power systems (HPS) such that the controlled system can enable the exchange between different power units \cite{5773511,5288888,zhou2011energy}. The concept of virtual power plant (VPP) was proposed as a cluster of distributed generation installations that is controlled by a central authority \cite{pudjianto2007virtual,you2009market,petersen2012optimal}. For example, generation sources with different dynamic response rates can be combined to achieve a higher operational efficiency that cannot be achieved by any of the sources used individually \cite{roy2010optimum,colson2011evaluating}. HPS can not only stabilize its output to have minimum oscillation, but it can be built to provide ancillary service. In contrast to generation management is flexible loads management that is optimized to utilize energy at the right time with the right price. Flexible demands can be classified according to their operational requirements. Non-interruptible loads, such as washing machines or dishwashers, have the lowest level of flexibility that cannot be interrupted in operation, whereas interruptible loads provide more degrees of freedom as long as consumer specified requirements are guaranteed to be met. Intra-microgrid load coordination can be extended to inter-microgrids coordination within the vicinity of the same distribution network. When each microgrid operator can extract the aggregated demand preferences representing the overall utility function, distributed or decentralized algorithms can be applied among microgrid operators and the regional transmission operators to reduce the costs of consuming electricity, the costs of transmission congestion, and the need for operating reserves \cite{li2011optimal,ilic2011efficient,6670130}.

\section{Smart Buildings}\label{smart building}
Smart buildings, both residential and commercial, are the most important elements in the microgrids that account for a surprising 40\% of total energy demand. The objective of building operation is to improve the reliability, sustainability, and efficiency of electricity usage while guaranteeing the comfort requirements of the building occupants. Thanks to the fast paced development of networked control technologies \cite{BaillieulAntsaklis2007} and novel concepts enabled by smart appliances, the control of a group of appliances has become a reality that enables the building to become smart.  Advanced buildings consumption analytics are provided by various companies, like OPower, C3 Energy, FirstFuel, to assist consumers in learning their energy consumption characteristics and help to improve energy efficiency. Companies like Honeywell, ABB, and Nest produce smart interfaces that have IP addresses to be controlled remotely from the Internet. The controlled appliances can be deployed in demand response programs that are executed to improve the economics of energy usage. The bar chart below shows the percentage of electricity consumption from different appliances in both residential and commercial buildings. It can be seen that loads with energy storage, such as HVAC, refrigeration, consume nearly 50\% of the overall consumption. Lighting in commercial buildings also provides big opportunity to affect the overall consumption if it can be operated at multiple dimming modes. In addition, non-interruptible but delay-able loads, such as dryers or washer, can contribute to the flexible shifting of aggregated consumption. 
\begin{figure}[hbt]
\centering
\includegraphics[width=8.4cm]{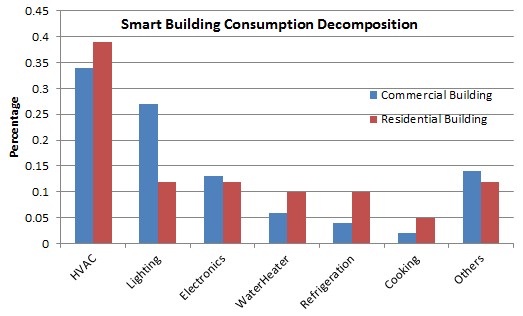}
\caption{Bar chart of electricity consumption in both commercial and residential buildings. Flexible loads, such as HVAC, refrigeration, and lighting, consume more than 50\% of the aggregated consumption.}
\label{building consumption}
\end{figure}

The first step towards smart building management is to build accurate models to describe the dynamics of the states that are controlled by the appliances. Early work focusing on the modelling of individual and aggregated HVAC appliance was proposed in \cite{IharaSchweppe1981,336078,4111301}. The thermodynamic model of the room temperature, which is a complex function determined by the building construction, thermal sources, and meteorological inputs, can be reduced to a linear differential equation with a few parameters containing the effective thermal time constant, temperature gain of appliances, etc. all of which are estimated by designing a cold load pick-up process. Following the idea of duty cycle modelling, controlled HVAC aggregated modelling in demand side management has been studied extensively in the past decade. The change of load diversity and synchronization was analysed in the case that direct load control (DLC) is used in demand reduction to shift peak load to non-peak hours \cite{LuChassin2004,LuChassin2005,BowenBaillieul2013_2,LeeKimKim2011}. When aggregated HVAC appliances are used to provide fast ramping regulation reserves, the load diversity will be affected by the past DLC profiles and price signals. This results in even more complex representation of the diversity \cite{Callaway2011-2,Callaway2012,ZhangCaramanisBaillieul,blum2014dynamic}, which is then characterized by a few statistical parameters that can be used as the state variable representing the aggregation.

Upon understanding the dynamical model of appliances, control architectures can be applied to provide demand reduction with minimum users' comfort degradation \cite{lee2011demand,BowenBaillieul2013_2}. Ancillary service can also be enabled to reduce the need for secondary reserve in the presence of renewables \cite{6760990,BowenBaillieul2013}. For example, the thermal dynamic model provides capacity bounds on the amount of regulation reserve that smart buildings can provide \cite{BowenBaillieul2013,6760776}. With a reasonable level of reserve capacity, the building operator can design either actuator based DLC or price based indirect load control to modulate the aggregated consumption \cite{callaway2011achieving,ramanathan2008framework,subramanian2012real,6152192,6160541}. If this was done with inaccurate thermal modeling and capacity overestimation, it could only achieve 50\% to 60\% of current demand response pilot programs \cite{inaccuratedr}. We formulate an optimization problem that can jointly minimize the summed costs of providing regulation reserve along with a consumer's disutility. The formulation is either based on dynamic system models or Markov decision process. Another source of flexibility is provided by interruptible loads that are deadline constrained, e.g. electric vehicles, washing machines. For example, large groups of electric vehicles can be coordinated to fill the overnight demand valley, to mitigate renewable energy intermittency, and to reduce transmission congestion \cite{lopes2011integration,gan2011optimal,ma2013decentralized,5399955,kempton2005vehicle,gu2010look,rotering2011optimal}.

In addition to demand response and the grid-wide benefits of ancillary services, smart building micro-grids have been identified as potential high-value resources for energy conservation and sustainability. There has been a standardization of an energy efficiency index that relates a smart building’s consumption to a reference level of consumption \cite{gonzalez2011towards,rietbergen2010setting,dubois2011energy}. Measurements of physical characteristics of consumption can be compared with estimates of activity levels of building occupants to support this index. Energy savings based on direct sensing of room occupancy and task-specific lighting, when combined with advances in bulb efficiency, can provide significant reductions in consumption, on the order of 44\% to 50\% according to some reports \cite{dubois2011energy,wang2014building,6653884}. 

\section{Energy Packets and PDLC}\label{energy packet and the pdlc}
The idea of an energy packet and its associated control protocols are motivated by the notion of data packets and packet switched communication networks. We see a lot of structural and functional similarities between communication and power systems -- they both have networked topology in which resources should be delivered with minimum disruption. Packet switched communication has shown its advantage in improving system fairness and network efficiency with limited transmission capacity. This is also a critical issue in power systems, namely how we can improve the energy quality of service to guarantee minimum energy disruption in demand response given a prescribed level of energy capacity. 

In this section we introduce \textit{energy packets}, as well as the associated control and communication protocols built on them that we call \textit{packetized direct load control} (PDLC). Differing from traditional energy distribution where appliances consume electricity according to their duty cycles, in PDLC appliances will consume quantized energy in the form of energy packets defined as follows:

\textit{Definition 1.} An \textit{energy packet} for a given appliance is a fixed time interval $\delta$ during which electricity is consumed at the appliance's rated power with its nominal voltage and current.
\begin{figure}[h]
\begin{center}
\includegraphics[width=7cm]{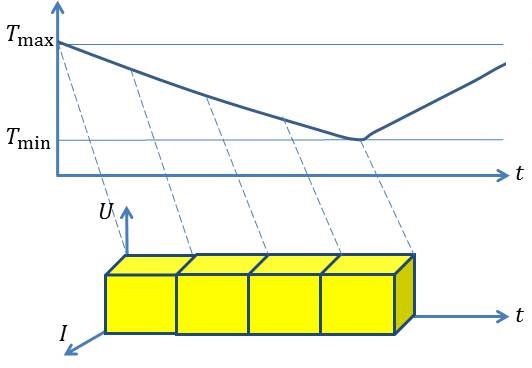}    
\caption{Continuous electricity consumption in the operating cycle is quantized into energy packets with short duration of $\delta$. Energy demand in a traditional duty cycle is therefore successive multiple demands of energy packets.} 
\label{fig: energy packet}
\end{center}
\end{figure}

Fig.~\ref{fig: energy packet} illustrates the concept of energy packets where continuous demand of electricity is quantized into multiple successive demands. The unit of demand and supply for electricity is therefore one energy packet. The value of $\delta$ can be designed based on the contract between the operator and consumers and will affect system performance. This will be discussed later.

In PDLC, group of flexible appliances of each consumer is connected to the central controller operated by the smart building operator (SBO) through a local area network where bi-directional communication/control is established such that: (i) appliances can send to the SBO their instantaneous state information signal that represents their desired comfort level or preferences of electricity consumption, and (ii) the SBO will send a binary control signal that executes direct control of the on/off switch of each appliance. In the discrete time control system, $\delta$ is both the duration of the energy packet and the decision interval. The communication from the appliances to the SBO indicates their energy needs for the next interval. Different choices of communication protocols can be built on top of the PDLC framework so that consumers can send various types of information based on their willingness to share information. An ideal communication signal from consumers contains the state of both the appliance and the environment regulated by the appliance. For example, an air conditioning appliance can report both its operating status as well as the desired temperature set point. As an alternative communication protocol, the appliances can choose to send binary information to the SBO stating whether or not they wish to consume a $\delta$-packet of electricity. This scenario corresponds to consumers wishing to protect their privacy or to a building automation monitoring system that is not reliable. The control from the SBO, which is executed at the beginning of each interval, is the decision on energy packets authorization to all appliances for the following one interval based on the received information. An appliance is allowed to consume one energy packet if it receives the authorization, and it needs to send signals in future intervals if additional packets are needed. The following few points comprise the remaining background of PDLC.

(1) The PDLC targets at controlling duty cycle appliances with thermal storage that includes air conditioners, water heaters, etc. Since our thermal models duty cycle appliances are all governed by first order ordinary differential equations, results in the rest of the paper focusing on the control of air conditioning units can be easily extended to alternative control objectives with minor changes.

(2) We assume duty cycle appliances have binary on/off operation modes. In the on mode, an appliance can only operate at its rated power and consume one unit of energy packet given the authorization from the SBO. In the off mode, an appliance consumes zero energy. There is no intermediate operation mode where consumers can flexibly adjust the power consumption of its appliance between the rated power and zero.

(3) Various types of appliances are separated by different feeders that are connected to the SBO. Appliances belonging to the same feeder have the same rated power and therefore consume the same unit of energy packets. The control of electricity consumption in the smart building is based on the parallel PDLC of each feeder. For inelastic and uncontrollable appliances, we assume they are connected to alternative feeders that are excluded from the analysis of the PDLC framework.

\section{Hierarchy of Information Communication}\label{Hierarchy of Information Communication}
We would like to characterize the best performance under two different levels of communication scenarios. The performance measure should cover both the system aspect and consumer aspect. For the system aspect, the controlled output from the SBO, which is the aggregated electricity consumption, would like to follow the purchased amount of energy packets with minimum deviation. For consumer aspect, the SBO should provide energy packets to guarantee that consumers' comfort be maintained within the desired range. In the following, we first discuss the baseline control performance where full state information is transmitted from appliances to the SBO. The second part will consider constrained communication where limited binary information is transmitted.

\subsection{Full Information Communication}
In the first communication scenario, we assume that sensors are installed around air conditioning units and real time data/information is sent to the SBO at a frequency higher than $\delta$ through an erasure free communication channel based on the local area network. At time $t$, the information transmitted from consumer $i$ is a paired value of temperature and operating status $\{T^{i}(t),0(1)\}$ where 0 (1) means appliance is at duty off (on) cycle. The objective of the SBO is to determine and reserve the minimum needed number of energy packets $m$ from the day ahead market to serve appliances within the building such that all consumers' room temperatures can be controlled within their preferred range. To quantify consumers' preferences of comfort, each consumer $i$ is only required to provide two values to specify his/her preference of comfort temperature control: a preferred set point value $T_{set}^{i}$ and a preferred maximum deviation value $\Delta^{i}$. The control system performance is therefore evaluated by $(i)$ consumption deviation away from the reserved $m$ energy packets and $(ii)$ consumers' dissatisfaction if the room temperature cannot be controlled within $[T_{set}^{i}-\Delta^{i},T_{set}^{i}+\Delta^{i}]$ at some time. 

Based on the real time $T(t)$ collected from all appliances, the SBO can predict the evolution of $T(t)$ for the next $\delta$ time interval based on the dynamic model \cite{IharaSchweppe1981}
\begin{equation}\label{thermal model}
\frac{d}{d t}T(t)=\frac{T_{out}-T(t)-T_{g}u+w(t)}{\tau},
\end{equation}
where $T_{out}$ is the outside temperature, $T_{g}$ is the temperature gain when the AC operates, $\tau$ is the effective thermal time constant, $w(t)$ is bounded measurement error or stochastic disturbance, and $u$ is a binary variable specifying the on/off operating status, as well as the instant temperature information of all consumers. This full information enables the SBO to: 1) allocate packets to pre-cool a room even if energy is not needed to remain in the comfort zone; 2) withdraw a packet from rooms having high temperature but still capable of acting as heat absorbing sources before temperatures reach the maximum comfort threshold. 

We show in \cite{BowenBaillieul2012} that the minimum number of energy packets $m$ that need to be purchased at any time $t$ can be determined by the total number of consumers $N$, that consumers' preferred set points, and properties of the appliance
\begin{equation}\label{critical amount of energy}
m=\frac{NT_{out}-\sum_{i=1}^{N}T_{set}^{i}}{T_{g}}.
\end{equation}
In addition, with proper choice of energy packet length $\delta$, the smart building can consume exactly $m$ energy packets at any decision interval such that no electricity consumption deviation would occur while maintaining the room temperature of all consumers to lie within their designated comfort bands. The result is formally stated as follows:

\textbf{Proposition 1} Assuming that $T^{i}(t^{\star})\in[T_{set}^{i}-\Delta^{i},T_{set}^{i}+\Delta^{i}]$ for all $i$ and the SBO purchases a fixed number $m=(NT_{out}-\sum_{i=1}^{N}T_{set}^{i})/T_{g}$ of packets, then there exists a positive $\delta$ such that with a fixed number of $m$ packets allocated starting from time $t^{\star}$, the PDLC can control all consumers' room temperatures to be in $[T_{set}^{i}-\Delta^{i},T_{set}^{i}+\Delta^{i}]$ for all $t\geq t^{\star}$.

To this end, it can be seen that $m$ in (\ref{critical amount of energy}) is the critical amount of energy that needs to be reserved in full information PDLC. The control performance based on full information is \textit{perfect} in the sense that no aggregated consumption deviation from $m$ and consumer's comfort loss occurs.

\subsection{Binary Information Communication}
In the binary communication setup, we fix the value of $m$ as in (\ref{critical amount of energy}) and derive the control performance for this protocol. When the SBO cannot acquire real time full information from appliances due to consumer privacy concerns, inaccurate temperature monitoring, etc, we consider a constrained communication that the appliances only signal the SBO binary information regarding their need of energy packets for the next interval. Namely the appliance will send a request signal to the SBO if it wishes to consume energy or a relinquish signal if it does not need energy for the next interval. The information loss to the SBO is the $T^{i}(t)$ that was used to construct the thermal dynamics in (\ref{thermal model}). The SBO, after receiving all requests during each interval, will authorize energy packets to a maximum number of $m$ for the next interval based on first-in-first-serve queuing principle. For requests received after the first $m$, the SBO will hold them and activate authorization in the future intervals. 

Since the SBO does not have real time temperature information, it models the energy request (arrival) and withdrawal (departure) process based on the duty cycle of appliances. Assuming the energy packet request rate from an idle AC is $\lambda$ (which is related to duty cycle off time as $1/\lambda$), the withdrawal rate from an operating AC is $\mu$ (which is related to duty cycle on time as $1/\mu$), and the energy packet length is $\delta$, the binary information based control system can be described as a closed loop queuing network with $N$ appliances and $m$ servers. As such, there is an associated probability distribution $\bar{p}(n,\delta)$ for the number of consuming appliances $n=0,\ldots,m$ in steady state. This distribution is determined by the steady state probability distribution of the number of appliances $p(x,\delta)$ for $x=0,\ldots,N$ in the queue.  

When $x$ appliances are in the queue at time $t$, $k=\min(x,m)$ servers are operating to serve the appliances with packet duration $\delta$. All $k$ appliances will finish the energy packet $[t,t+\delta]$ by the end of the interval and they will independently decide whether to request additional packets. With probability $p(\delta)=e^{-\mu\delta}$, an appliance will request an additional packet and there is no departure in this case. With probability $1-p(\delta)=1-e^{-\mu\delta}$, an appliance will switch into the idle state that results in one departure. Since the probability of departure linearly increases with the departure rate for small value of $\delta$, the departure rate for an appliance is $(1-p(\delta))/\delta$. Therefore the departure rate with $k$ operating appliances is $k[1-p(\delta)]/\delta$. In addition, the arrival rate that packet requests are received from idle appliances is $(N-x)\lambda$. We can solve for the steady state probability distribution $p(x,\delta)$ based on the departure/arrival rate
\begin{displaymath}
p(x,\delta)(N-x)\lambda = p(x+1,\delta)k[1-p(\delta)]/\delta,
\end{displaymath}
where $k=\min(x,m)$. This yields
\begin{displaymath}
p(x,\delta)=\bigg\{
\begin{array}{l}
p(0,\delta)\binom{N}{x}r(\delta)^{x},\hspace{3mm} x<m \vspace{2mm}\\
p(0,\delta)\binom{N}{x}r(\delta)^{x}\frac{\displaystyle x!}{\displaystyle m^{x-m}m!},\hspace{3mm} x\geq m,
\end{array}
\end{displaymath}
where $r(\delta) = \frac{\displaystyle\lambda\delta}{\displaystyle 1-e^{-\mu\delta}}$. The steady state probability distribution can be gotten after normalizing $p(x,\delta)$.

$p(x,\delta)$ can be used to quantify the performance of the closed loop queuing network that includes the mean queue length, the average service (waiting pluses operating) time, etc. We define the number of operating appliances as 
\begin{equation} \label{eq: pi}
\bar{p}(n,\delta) = \bigg\{
\begin{array}{l}
 p(n,\delta), \hspace{2mm} n=0,\ldots,m-1\\
\sum\limits_{k=m}^{N} p(k,\delta), \hspace{2mm} n=m
\end{array}.
\end{equation}
Then the first component of the system performance, which is the uncertainty of energy packet consumption is the variance of the random variable $n$ based on the distribution in (\ref{eq: pi}). The measure of consumers' comfort loss can be calculated based on the total waiting time for consumers to complete their energy needs. The average number of consumers in the queue is 
\begin{displaymath}
Q(m,\delta)=\sum\limits_{x=0}^{N}p(x,\delta)x.
\end{displaymath}
The average arrival rate is 
\begin{displaymath}
\lambda_{\textrm{ave}}=\lambda(N-Q(m,\delta)).
\end{displaymath}
Based on Little's Law, the average time consumers spent in the queue, namely the service time, is
\begin{displaymath}
S(m,\delta)=\frac{Q(m,\delta)}{\lambda(N-Q(m,\delta))}.
\end{displaymath}
Since the uncontrolled duty cycle expected operating time is $1/\mu$, the extra time that consumer's spent in the system is
\begin{equation}\label{extra time in the system}
W(m,\delta)=S(m,\delta)-\frac{1}{\mu}.
\end{equation}
$W(m,\delta)$ characterizes the binary information based control performance for consumer utility. The system measure of performance is the uncertainty (variance) of the distribution $\textrm{Var}(\bar{p}(n,\delta))$. We showed in \cite{zhang2014communication} that there is a trade off between $W(m,\delta)$ and $\textrm{Var}(\bar{p}(n,\delta))$ as $\delta$ is varied.

\textbf{Proposition 2} As $\delta$ increases, $\textrm{Var}(n(m,\delta))$ will decrease and $W(m,\delta)$ will increase.

\textbf{Remark 1.} Proposition 1 and Proposition 2 provide the quantification of the difference in system performance when information is degraded from full to binary level. We see costs exist for both consumers (extra waiting time $W(m,\delta)$) and the overall system (demand uncertainty $\textrm{Var}(\bar{p}(n,\delta))$). In the following, we will transform these two quantifications into common measuring units and propose a corresponding energy reservation strategy for the optimal value of $m$.

\section{Metrics of System Performance}\label{Quantifying Performance Degradation}
When procuring the same energy packet level at $m$, Section \ref{Hierarchy of Information Communication} shows that both performance in terms of consumption uncertainty and consumers' dis-utility will degrade when the information level is degraded from continuous to binary. It is clear that procuring $m$ energy packets is optimal for the SBO having continuous state information. We next find the optimal energy procurement strategy when the SBO only has access to binary information. In what follows, we will provide two metrics that will be used in optimizing $m$. For notational simplicity, the packet duration $\delta$ will be fixed based on the contract established between the SBO and consumers and will be omitted.

\subsection{Energy Function Metric}
The energy metric optimization is based on the intuition that a large number of reserved energy packets will result in unnecessary excess capacity, while a small number of energy packets will result in undesirable capacity deficiency. The optimal decision must strike the balance between the two costs. Based on the steady state probability distribution of consumers in the queue $p(x)$ for the system procuring $m$ energy packets in the day ahead market, we define the {\it capacity deficiency} as
\begin{displaymath}
De(m) = \sum\limits_{x=m+1}^{N} p(x)(x-m).
\end{displaymath}
And we define the {\it excess capacity} as
\begin{displaymath}
Ex(m) = \sum\limits_{x=0}^{m-1} p(x)(m-x).
\end{displaymath}
Since the units of both the excess capacity and the capacity deficiency are energy packets, we propose that the optimal energy packet procurement $m^{\star}$ can be gotten by minimizing the summed cost of the two metrics. Denote $T(m)$ as the total performance degradation, the SBO can determine the optimal energy procurement $m^{\star}$ based on energy metric minimization  
\begin{equation}\label{optimal energy reserve}
\min_{m\in[1,N]} \textbf{\textrm{E}}(m)=Ex(m) + De(m). 
\end{equation}
We show that $\textbf{\textrm{E}}(m)$ is a convex function guaranteeing that the optimal $m^{\star}$ is unique. 

\textbf{Proposition 3} $\textbf{\textrm{E}}(m)$ is a convex function of $m$.

\textit{Proof.} We prove by showing that both the excess capacity $Ex(m)$ and the capacity deficiency $De(m)$ are convex functions of $m$. The relation between the excess capacity $Ex(m)$ and the system throughput $Th(m)$ is
\begin{displaymath}
[m-Ex(m)]\mu= Th(m).
\end{displaymath}
It has been shown that in a closed queuing network $Th(m)$ is a concave function of $m$ \cite{shanthikumar1989second}, and therefore $Ex(m)$ is a convex function of $m$. To see the convexity of $De(m)$, we have
\begin{equation}\label{capacity deficiency}
\begin{array}{lll}
De(m) & = & \sum\limits_{x=m+1}^{N} p(x)(x-m)\\
& = & \sum\limits_{x=0}^{N} p(x)(x-m) - \sum\limits_{x=0}^{m} p(x)(x-m) \\
& = & Q(m) - m + Ex(m)
\end{array}
\end{equation}
where $Q(m)$ is the mean number of consumers in the queue. The steady state arrival rate of the closed queue is product of the single arrival rate and the mean number of idle appliances: $\lambda[N-Q(m)]$. The steady state departure rate is the throughput of the system: $\mu[m-Ex(m)]$. Since the steady state arrival and departure rates are equal to each other
\begin{equation}\label{steady state throughput}
\lambda[N-Q(m)]=\mu[m-Ex(m)],
\end{equation}
we have
\begin{equation}\label{mean queue length}
Q(m) = N - \frac{\mu}{\lambda}[m-Ex(m)].
\end{equation}
Substituting (\ref{mean queue length}) into (\ref{capacity deficiency}) we get
\begin{displaymath}
De(m) = N-(1+\frac{\mu}{\lambda})[m-Ex(m)]
\end{displaymath}
is a convex function of $m$. \hfill\(\Box\)

If we express both $Ex(m)$ and $De(m)$ as a function of the mean queue length $Q(m)$, then $\textbf{\textrm{E}}(m)$ becomes 
\begin{equation}\label{tm as a function of qm}
\textbf{\textrm{E}}(m)=(1+2\lambda/\mu)Q(m)+m-2\lambda/\mu N.
\end{equation}
Hence the optimal $m^{\star}$ is chosen to satisfy 
\begin{displaymath}
Q'(m^{\star})=-\frac{1}{1+2\lambda/\mu}.
\end{displaymath}

\textbf{Remark 2.} The energy metric optimization is not restricted to the form in (\ref{optimal energy reserve}). In fact we can add weight coefficient on both $Ex(m)$ and $De(m)$. One reasonable weight on $Ex(m)$ is the day-ahead energy costs penalizing unnecessary procurement, and weight on $De(m)$ is the predicted real time energy costs penalizing the balancing purchase caused by deficiency.

\subsection{Welfare Function Metric}
The concept of capacity deficiency and excess capacity can be translated to derive a welfare metric based optimization -- (i) capacity deficiency results in a dis-continuity of electricity consumption and therefore results in room temperature deviations from the allowable range, and (ii) excess capacity incurs unnecessary energy reserve costs. These two costs are quantified as follows.

The capacity deficiency characterizes the extra waiting time consumers need to spend in the queuing system to complete the energy packets service for one duty cycle. Based on (\ref{capacity deficiency}) and (\ref{steady state throughput}), $De(m)$ is related to $Q(m)$ by
\begin{displaymath}
De(m) = (1+\frac{\lambda}{\mu})Q(m) - \frac{\lambda}{\mu}N.
\end{displaymath}
Therefore the consumer's extra time spent in the system, $W(m)$ in (\ref{extra time in the system}), can be derived from Little's Law:
\begin{displaymath}
W(m)=\frac{De(m)+\frac{\lambda}{\mu}N}{\lambda[N-De(m)]}-\frac{1}{\mu}.
\end{displaymath}
It can be shown that $W'(m)<0$, $W''(m)>0$, and therefore $W(m)$ is a decreasing convex function. When the comfort bandwidth is relatively small compared to the distance between indoor and outdoor temperatures, the room temperature deviation $\Delta T(m)$ drifts linearly with the consumers' waiting time. Hence
\begin{displaymath}
\Delta T(m) = \kappa W(m),
\end{displaymath}
where temperature drift rate $\kappa$ can be determined by the duty off cycle. The welfare characterization of consumer's dis-utility $g(\cdot)$ can be defined as a function of the temperature deviation $\Delta T(m)$ as $g(\Delta(m))$. If the utility function $g(\Delta T(m))$ is convex and non-decreasing w.r.t $\Delta T(m)$, as assumed in \cite{Hammerstrom2007}, \cite{BilginCaramanis2012}, $g(\Delta T(m))$ is then a monotonically decreasing and convex function. This can be easily checked by the first and second order conditions
\begin{displaymath}
\frac{d}{dm}g(\Delta T(m))=\frac{d}{d \Delta T(m)}g(\Delta T(m))\Delta T'(m)<0,
\end{displaymath}
and
\begin{displaymath}
\begin{array}{lll}
\frac{d}{dm^{2}}g(\Delta T(m)) & = & \frac{d}{d \Delta T(m)^{2}}g(\Delta T(m))\Delta T'(m)^{2} + \\
&& \frac{d}{d \Delta T(m)}g(\Delta T(m))\Delta T''(m)>0.
\end{array}
\end{displaymath}
Hence $g(\Delta T(m))$ will be used to translate capacity deficiency in energy metric optimization into consumers' utility in welfare optimization. 

Next we translate excess capacity into the welfare metric which can be done straightforwardly as follows. The penalty $h(\cdot)$ for having of excess capacity can be defined as a linear function of $Ex(m)$ where the coefficient can be either the energy purchasing price or the energy reserving price depending on the contract between the SBO and the ISO. Given $Ex(m)$ is convex, $h(\cdot)$ is also convex. To this end, the monetary metric optimization solves for $m^{\star}$ such that
\begin{displaymath}
m^{\star}= \arg \min_{m\in[1,N]} g(\Delta T(m)) + h(Ex(m)).
\end{displaymath}
Similar to the energy metric optimization, we can transform both $\Delta T(m)$ and $Ex(m)$ to $Q(m)$. The unique welfare based solution can be gotten by minimizing the welfare metric $\textbf{\textrm{W}}(m)$ defined as
\begin{equation}\label{monetary optimization}
\textbf{\textrm{W}}(m) = g\big(\kappa(\frac{Q(m)}{\lambda(N-Q(m))}-\frac{1}{\mu})\big) + h\big(m-\frac{\lambda}{\mu}[N-Q(m)]\big).
\end{equation}

\section{Impact of Renewable Energy}\label{Impact of Volatile Resources}
When the SBO procures day-ahead energy packets from traditional resources, there is no volatility in terms of the actual amount of energy packets that the SBO can get. Therefore the day-ahead energy procurement strategy is to purchase $m^{\star}$ minimizing the energy metric (\ref{optimal energy reserve}) or the welfare perspective (\ref{monetary optimization}). However, when the SBO purchases energy with a combination of deterministic (traditional) and volatile (renewable) resources, the actual number of available packets as well as the real time performance will be stochastic that depends on the probability distribution of the volatile resources. In the following, we analyze the impact of volatile energy on the system performance. 

We begin by modeling the distribution of volatile resources and focus on wind generation. We use a Gaussian as the wind prediction error model. This is typically used for short term wind prediction for time scales between 6 and 48 hours ahead of operating time \cite{hodge2011wind}, \cite{4530750}. Suppose the SBO has access to wind generation resources, and suppose the realization of wind energy $P_{v}$ satisfies the following Gaussian distribution with prediction mean $P_{r}$ and variance $\sigma^{2}$
\begin{displaymath}
P_{v} \sim N(P_{r},\sigma^{2}).
\end{displaymath}
In addition to wind energy, the SBO can choose to purchase a certain amount of traditional energy denoted as $P_{t}$. This yields the following probability distribution of the number of packets $m$ that will be available in real time 
\begin{equation}\label{normal distribution of m}
m = P_{v} + P_{t} \sim N(P_{r} + P_{t}, \sigma^{2}),
\end{equation}
and yields the following expected system welfare $\bar{\textbf{\textrm{W}}}(P_{r},P_{t},\sigma)$ defined in (\ref{monetary optimization})
\begin{equation}\label{expected costs with renewable}
\bar{\textbf{\textrm{W}}}(P_{r},P_{t},\sigma)=\int\limits p(m,P_{r},P_{t},\sigma)\textbf{\textrm{W}}(m)dm
\end{equation}
where $p(m,P_{r},P_{t},\sigma)$ is the probability that the SBO will have $m$ number of available packets governed by the Gaussian distribution in (\ref{normal distribution of m}). Similar to the uniqueness of $m^{\star}$ when the SBO procures solely traditional energy, we prove that the optimal procurement $P_{t}$ is unique when the SBO is provided with the wind resources distribution.

\textbf{Proposition 4} For given $\{P_{r},\sigma\}$ characterizing the wind energy, there is a unique $P_{t}$ that minimizes $\bar{\textbf{\textrm{W}}}(P_{r},P_{t},\sigma)$.

\textit{Proof.} Since $m \sim N(P_{t}+P_{r},\sigma^{2})$, we have

\begin{displaymath}
\bar{\textbf{\textrm{W}}}(P_{r},P_{t},\sigma) = \int\limits \frac{1}{\sqrt{2\pi\sigma^{2}}}
e^{-\frac{(m-P_{r}-P_{t})^{2}}{2\sigma^{2}}}  \textbf{\textrm{W}}(m)dm.
\end{displaymath}
Letting $M=(m-P_{r}-P_{t})/\sigma$ be a change of variable
\begin{equation}\label{total cost proposition 2}
\bar{\textbf{\textrm{W}}}(P_{r},P_{t}) = \frac{1}{\sqrt{2\pi\sigma^{2}}} \int\limits e^{-\frac{M^{2}}{2}}  \textbf{\textrm{W}}(P_{r}+P_{t}+M\sigma)dM.
\end{equation}
Based on Leibniz integral rule, the second order derivative of the cost function w.r.t $P_{t}$ is
\begin{equation}\label{derivative}
\frac{1}{\sqrt{2\pi\sigma^{2}}} \int\limits e^{-\frac{M^{2}}{2}} \frac{\partial}{\partial P_{t}^{2}}\textbf{\textrm{W}}(P_{r}+P_{t}+M\sigma)dM,
\end{equation}
which is strictly positive due to the convexity of $\textbf{\textrm{W}}(\cdot)$. Hence $\bar{\textbf{\textrm{W}}}(P_{r},P_{t},\sigma)$ is a convex function of $P_{t}$ for volatile features $\{P_{r},\sigma\}$, and we have a unique optimal $P_{t}$. \hfill\(\Box\)

We proceed to analyze the impact of wind energy on the optimal system welfare. We first discuss the scenario when $P_{r}$ and $\sigma$ are uncorrelated. The result is shown in Proposition 3 that increased uncertainty $\sigma$ will decrease the welfare. This result is then extended to wind output having correlated mean and variance where the variance linearly increases with the mean $\sigma=kP_{r}$. This corresponds to the case where the wind farm output is the summation of small wind mills locating at the same spot.

\textbf{Proposition 5} Define
\begin{equation} \label{optimal system welfare}
\textbf{\textrm{F}}(P_{r},\sigma)=\min_{P_{t}} \bar{\textbf{\textrm{W}}}(P_{r},P_{t},\sigma) 
\end{equation} 
then $F(\cdot,\cdot)$ increases as $\sigma$ increases.

\textit{Proof.} Consider two levels of volatile resource uncertainty $\sigma_{1}<\sigma_{2}$. Denote $P_{t_{2}}$ as the optimal solution for $\{P_{r},\sigma_{2}\}$, and denote $\bar{P}=P_{r}+P_{t_{2}}$ as the mean number of packets expected in real time. We compare the value of $\bar{\textbf{\textrm{W}}}(\bar{P},\sigma_{1})$ and $\bar{\textbf{\textrm{W}}}(\bar{P},\sigma_{2})$. 

The probability distribution of the number of packets in the two scenarios are $m_{1}\sim N(\bar{P},\sigma_{1}^{2})$ and $m_{2}\sim N(\bar{P},\sigma_{2}^{2})$. Based on the symmetric property of Gaussian distributions, there is a positive value of $k$ such that 
\begin{displaymath}
\left\{ \begin{array}{ll}
p(m,\bar{P},\sigma_{1})\leq p(m,\bar{P},\sigma_{2}), & \textrm{if} \hspace{3mm} |m-\bar{P}|\geq k\\
p(m,\bar{P},\sigma_{1})>p(m,\bar{P},\sigma_{2}), & \textrm{otherwise}.
\end{array}\right.
\end{displaymath}
We have
\begin{equation}\label{uncertainty 1 equation}
\begin{array}{ll}
& \bar{\textbf{\textrm{W}}}(\bar{P},\sigma_{1})\\
=& \int\limits_{-\infty}^{\bar{P}-k}p(m,\bar{P},\sigma_{1})\textbf{\textrm{W}}(m)dm + \int\limits_{\bar{P}+k}^{+\infty}p(m,\bar{P},\sigma_{1})\textbf{\textrm{W}}(m)dm\\
& \int\limits^{\bar{P}}_{\bar{P}-k}p(m,\bar{P},\sigma_{1})\textbf{\textrm{W}}(m)dm + \int\limits_{\bar{P}}^{\bar{P}+k}p(m,\bar{P},\sigma_{1})\textbf{\textrm{W}}(m)dm.
\end{array}
\end{equation}
Letting $m=2\bar{P}-M$ and using the fact that a Gaussian distribution is symmetric around $\bar{P}$: $p(m,\bar{P},\sigma_{1})=p(2\bar{P}-m,\bar{P},\sigma_{1})$ we have
\begin{equation}\label{equivalence1}
\begin{array}{lll}
&&\int\limits_{\bar{P}+k}^{+\infty}p(m,\bar{P},\sigma_{1})\textbf{\textrm{W}}(m)dm\\
&=&\int\limits_{\bar{P}-k}^{-\infty}p(2\bar{P}-M,\bar{P},\sigma_{1})\textbf{\textrm{W}}(2\bar{P}-M)d(2\bar{P}-M),\\
&=&\int\limits^{\bar{P}-k}_{-\infty}p(M,\bar{P},\sigma_{1})\textbf{\textrm{W}}(2\bar{P}-M)dM.
\end{array}
\end{equation}
Similarly,
\begin{equation}\label{equivalence2}
\int\limits^{\bar{P}+k}_{\bar{P}}p(m,\bar{P},\sigma_{1})\textbf{\textrm{W}}(m)dm=\int\limits_{\bar{P}-k}^{\bar{P}}p(m,\bar{P},\sigma_{1})\textbf{\textrm{W}}(2\bar{P}-m)dm.
\end{equation}
Substituting (\ref{equivalence1}) and (\ref{equivalence2}) into (\ref{uncertainty 1 equation}) we have
\begin{equation}\label{expression for sigma 1}
\begin{array}{ll}
\bar{\textbf{\textrm{W}}}(\bar{P},\sigma_{1})=& \int\limits^{\bar{P}-k}_{-\infty}p(m,\bar{P},\sigma_{1})[\textbf{\textrm{W}}(m)+\textbf{\textrm{W}}(2\bar{P}-m)]dm+\\
& \int\limits_{\bar{P}-k}^{\bar{P}}p(m,\bar{P},\sigma_{1})[\textbf{\textrm{W}}(m)+\textbf{\textrm{W}}(2\bar{P}-m)]dm.
\end{array}
\end{equation}
Similarly we can have the expression of $\textbf{\textrm{W}}(\bar{P},\sigma_{2})$ which together with (\ref{expression for sigma 1}) is used to derive
\small
\begin{equation}\label{two procurement strategy}
\begin{array}{ll}
& \bar{\textbf{\textrm{W}}}(\bar{P},\sigma_{1})-\bar{\textbf{\textrm{W}}}(\bar{P},\sigma_{2})\\
=& \int\limits_{-\infty}^{\bar{P}-k}[p(m,\bar{P},\sigma_{1})-p(m,\bar{P},\sigma_{2})][\textbf{\textrm{W}}(m)+\textbf{\textrm{W}}(2\bar{P}-m)]dm + \\
& \int\limits_{\bar{P}-k}^{\bar{P}}[p(m,\bar{P},\sigma_{1})-p(m,\bar{P},\sigma_{2})[\textbf{\textrm{W}}(m)+\textbf{\textrm{W}}(2\bar{P}-m)]dm.
\end{array}
\end{equation}
\normalsize
For any $m_{1},m_{2}$ such that $m_{1}\leq \bar{P}-k\leq m_{2}\leq\bar{P}$, we have

\small
\begingroup
\addtolength{\jot}{1em}
\begin{eqnarray}
& [\textbf{\textrm{W}}(m_{1})+\textbf{\textrm{W}}(2\bar{P}-m_{1})]-[\textbf{\textrm{W}}(m_{2})+\textbf{\textrm{W}}(2\bar{P}-m_{2})]\nonumber\\
=& [\textbf{\textrm{W}}(m_{1})-\textbf{\textrm{W}}(m_{2})] + [\textbf{\textrm{W}}(2\bar{P}-m_{1})-\textbf{\textrm{W}}(2\bar{P}-m_{2})] \nonumber\\
\geq & \frac{m_{2}-m_{1}}{\bar{P}-m_{1}}[\textbf{\textrm{W}}(m_{1})-\textbf{\textrm{W}}(\bar{P})] + \frac{m_{1}-m_{2}}{m_{1}-\bar{P}}[\textbf{\textrm{W}}(2\bar{P}-m_{1})-\textbf{\textrm{W}}(\bar{P})] \nonumber\\
=& \frac{m_{2}-m_{1}}{\bar{P}-m_{1}}[\textbf{\textrm{W}}(m_{1})-2\textbf{\textrm{W}}(\bar{P})+\textbf{\textrm{W}}(2\bar{P}-m_{1})] \geq 0. \label{inequality}
\end{eqnarray}
\endgroup

\normalsize
The inequality is derived based on the Jensen's inequality on $\textbf{\textrm{W}}(\cdot)$. Let $C_{1} = \inf \Big\{\textbf{\textrm{W}}(m)+\textbf{\textrm{W}}(2\bar{P}-m) | m\in (-\infty,\bar{P}-k] \Big\}$ and $C_{2} = \sup \Big\{\textbf{\textrm{W}}(m)+\textbf{\textrm{W}}(2\bar{P}-m) | m\in [\bar{P-k,\bar{P}}] \Big\}$. From (\ref{inequality}) we have $C_{1} \geq C_{2}$. Therefore we have for (\ref{two procurement strategy})
\begin{displaymath}
\begin{array}{lll}
&& \bar{\textbf{\textrm{W}}}(\bar{P},\sigma_{1})-\bar{\textbf{\textrm{W}}}(\bar{P},\sigma_{2})\\
& \leq & C_{1} \int\limits_{-\infty}^{\bar{P}-k} p(m,\bar{P},\sigma_{1})-p(m,\bar{P},\sigma_{2})dm + \\
&& C_{2} \int\limits_{\bar{P}-k}^{\bar{P}} p(m,\bar{P},\sigma_{1})-p(m,\bar{P},\sigma_{2})dm, \\ 
& = & (C_{1}-C_{2}) \int\limits_{-\infty}^{\bar{P}-k}p(m,\bar{P},\sigma_{1})-p(m,\bar{P},\sigma_{2})dm \leq 0.
\end{array}
\end{displaymath}
The first inequality is derived since $p(m,\bar{P},\sigma_{1})<p(m,\bar{P},\sigma_{2})$ for $m\in (-\infty,\bar{P}-k)$ and $p(m,\bar{P},\sigma_{1})>p(m,\bar{P},\sigma_{2})$ for $m\in (\bar{P}-k,\bar{P})$. The second equality is derived from the property of Gaussian distributions
\begin{displaymath}
\int\limits_{-\infty}^{\bar{P}}\Big[p(m,\bar{P},\sigma_{1}) - p(m,\bar{P}\sigma_{2})\Big]dm = 0.
\end{displaymath}
Hence the system costs will increase when the uncertainty increases from $\sigma_{1}$ to $\sigma_{2}$
\begin{equation}\label{same mean diffrent variance}
\bar{\textbf{\textrm{W}}}(P_{r},P_{t_{2}},\sigma_{1}) \leq \bar{\textbf{\textrm{W}}}(P_{r},P_{t_{2}},\sigma_{2}).
\end{equation} 
Denoting $P_{t_{1}}$ as the minimizer for wind uncertainty level $\sigma_{1}$, we have
\begin{displaymath}
\begin{array}{l}
\bar{\textbf{\textrm{W}}}(P_{r},P_{t_{1}},\sigma_{1}) \leq \bar{\textbf{\textrm{W}}}(P_{r},P_{t_{2}},\sigma_{1}) \leq \bar{\textbf{\textrm{W}}}(P_{r},P_{t_{2}},\sigma_{2}).
\end{array}
\end{displaymath}
This ends the proof. \hfill\(\Box\)

We extend the analysis to the scenario when the wind output mean and variance correlate. Suppose a wind farm is composed of many wind turbines having relatively small value of output with Gaussian distributions 
\begin{displaymath}
\delta_{r} \sim N(\delta_{m},\sigma_{\delta}^{2}).
\end{displaymath}
Then the scaled output of the wind farm with $n$ wind turbines is
\begin{displaymath}
n\delta_{r} \sim N(n\delta_{m},(n\sigma_{\delta})^{2}),
\end{displaymath}
Clearly the standard deviation of the aggregated output $\sigma = n\sigma_{\delta}$ scales with mean output $P_{r} = n\delta_{m}$ with constant coefficient of variation $k=\sigma_{\delta}/\delta_{m}$. The more expected resources the SBO wishes to reserve, the more uncertainty it will face, and hence the system costs. This is formally stated as follows.

\textbf{Corollary 1} If the wind output has correlated mean and variance satisfying
\begin{displaymath}
P_{v}\sim N(P_{r},(kP_{r})^{2}),
\end{displaymath} 
then the system costs $\textbf{\textrm{F}}(P_{r},kP_{r})$ in (\ref{optimal system welfare}) will increase as $P_{r}$ increases.

\textit{Proof.} Consider two levels of wind output $P_{r_{1}}<P_{r_{2}}$, and denote $P_{t_{2}}$ as the optimal traditional energy procurement under $P_{r_{2}}$. From (\ref{same mean diffrent variance}) we know that we can carefully choose $P_{t}=P_{t_{2}}+P_{r_{2}}-P_{r_{1}}$ such that $\bar{\textbf{\textrm{W}}}(P_{r_{1}},kP_{r_{1}},P_{t})\leq \bar{\textbf{\textrm{W}}}(P_{r_{2}},kP_{r_{2}},P_{t_{2}})$. Denoting $P_{t_{1}}$ as the optimal solution under $P_{r_{1}}$, we will have
\begin{displaymath}
\bar{\textbf{\textrm{W}}}(P_{r_{1}},kP_{r_{1}},P_{t_{1}})\leq \bar{\textbf{\textrm{W}}}(P_{r_{1}},kP_{r_{1}},P_{t})\leq \bar{\textbf{\textrm{W}}}(P_{r_{2}},kP_{r_{2}},P_{t_{2}}).
\end{displaymath}
This ends the proof. \hfill\(\Box\)

Proposition 5 and Corollary 1 indicate that the system welfare cost will increase as either the mean or the variance of wind resources increases. The introduction of wind resources does not bring benefits to real time operation. However, wind energy is helpful from the day-ahead point of view by reducing the costs of purchasing $P_{t}$ that is needed otherwise. This yields a trade-off where we need to jointly minimize the day ahead energy procurement and real time operating welfare costs. In the following section, we discuss the optimal energy purchase strategy in both day-ahead and real time markets.

\section{Optimal Energy Procurement} \label{Optimal Hourly Energy Procurement}
We discuss the optimal energy purchasing strategy for the SBO in different market settings. We begin by focusing on single market participation in Section \ref{single market participation} where the SBO only participates in the day-ahead energy market by purchasing certain amounts of traditional and wind energy at given day ahead prices. On the next day, it will utilize what has been purchased. This analysis is then extended in \ref{double market participation} when the SBO can participate in both day-ahead and real time market. The SBO exersices the same choices in the day-ahead market as before. In addition, the SBO can choose to procure additional balancing energy or sell day-ahead purchased energy back to the market based on the realization of wind output. 

\subsection{Single Market Participation}\label{single market participation}
Suppose the SBO participates only in the day-ahead market and plans to optimally procure a certain number of energy packets from both traditional and volatile resources such that it minimizes the summed costs of the energy purchase and real time welfare costs $\bar{\textbf{\textrm{W}}}(P_{t},P_{r},\sigma)$. Here $\bar{\textbf{\textrm{W}}}(P_{t},P_{r},\sigma)$ is defined by (\ref{expected costs with renewable}) with $\textbf{\textrm{W}}(m)=g\big(\kappa(\frac{Q(m)}{\lambda(N-Q(m))}-\frac{1}{\mu})\big)$ containing only the consumer's waiting time to access energy packets. In the day ahead, the unit traditional energy price is known to the SBO as $k_{t}$. We consider the following two statistical characterizations of wind energy availability.

The first model characterises wind availability as a process with a fixed mean $P_{r}$ and standard deviation $\sigma$ that cannot be chosen by the SBO. For instance, the SBO can locally connect to a wind farm within its microgrid whose statistical output is determined by its location, time, etc. The SBO's objective is to minimize over $P_{t}$ while having fixed $P_{r}$ and $\sigma$:
\begin{equation}\label{only day ahead purchase of Pt}
\underset{P_{t}}{\text{min}} \hspace{3mm}k_{t}P_{t} + \bar{\textbf{\textrm{W}}}(P_{r},P_{t},\sigma).
\end{equation}
Clearly (\ref{only day ahead purchase of Pt}) is a convex function given the second term is convex as shown in Prop. 4. It is straightforward to find the unique optimal $P_{t}$.

In the second model the wind output has correlated mean and variance due to the aggregation of many small windmills. The wind output $P_{v}$ has the same distribution as in Corollary 1
\begin{displaymath}
P_{v}\sim N(P_{r},(kP_{r})^{2}).
\end{displaymath} 
The SBO can flexibly sign contracts for outputs of certain number of wind turbines in order to get wind energy in real time. The SBO would solve over $P_{t}$ and $P_{r}$ to minimize the total operation costs. (We omit the third argument in $\bar{\textbf{\textrm{W}}}(\cdot)$ since the variance depends on the mean.)
\begin{equation}\label{day ahead purchase of Pt Pr}
\underset{P_{t},P_{r}}{\text{min}} \hspace{3mm}k_{t}P_{t} + k_{r}P_{r} + \bar{\textbf{\textrm{W}}}(P_{r},P_{t}).
\end{equation}
In (\ref{day ahead purchase of Pt Pr})$, k_{r}$ is the unit reservation price for wind energy (satisfying $k_{r}<k_{t}$) that is used to cover the low operation costs and capacity costs of wind turbines. Proposition 6 proves that $\bar{\textbf{\textrm{W}}}(P_{t},P_{r},kP_{r})$ is jointly convex in $\{P_{t},P_{r}\}$, and this guarantees a unique procurement solution.

\textbf{Proposition 6} $\bar{\textbf{\textrm{W}}}(P_{t},P_{r})$ is jointly convex of $P_{t},P_{r}$.

\textit{Proof.} From (\ref{tm as a function of qm}) we have
\begin{displaymath}
\begin{array}{ll}
\bar{\textbf{\textrm{W}}}(P_{t},P_{r})&=\int\limits_{-\infty}^{\infty}p(m,P_{t},P_{r})\textbf{\textrm{W}}(m)dm,\\
& = \int\limits_{-\infty}^{\infty}\frac{1}{\sqrt{2\pi k^{2}P_{r}^{2}}}e^{\frac{(m-P_{t}-P_{r})^{2}}{2(kP_{r})^{2}}}\textbf{\textrm{W}}(m)dm,\\
& = \frac{1}{\sqrt{2\pi}}\int\limits_{-\infty}^{\infty}e^{-\frac{M^{2}}{2}}\textbf{\textrm{W}}(P_{t}+(1+kM)P_{r})dM.
\end{array}
\end{displaymath}
The last equation is gotten by a change of variable $M=\frac{m-P_{t}-P_{r}}{kP_{r}}$.
We prove the Hessian of $\bar{\textbf{\textrm{W}}}(P_{t},P_{r})$ is positive definite
\begin{displaymath}
\textit{\textbf{H}}(\bar{\textbf{\textrm{W}}})(P_{t},P_{r})=\left[ \begin{array}{cc}
\frac{\partial^{2}}{\partial P_{t}^{2}}\bar{\textbf{\textrm{W}}} & \frac{\partial^{2}}{\partial P_{t} \partial P_{r} }\bar{\textbf{\textrm{W}}}\\
\frac{\partial^{2}}{\partial P_{r} \partial P_{t} }\bar{\textbf{\textrm{W}}} & \frac{\partial^{2}}{\partial P_{r}^{2}}\bar{\textbf{\textrm{W}}}
\end{array} \right] \succeq 0.
\end{displaymath}
Based on Leibniz integral rule
\small
\begin{equation}\label{partial deravitives of pr pt}
\begin{array}{l}
\frac{\partial^{2}}{\partial P_{t}^{2}}\bar{\textbf{\textrm{W}}} = \frac{1}{\sqrt{2\pi}}\int\limits_{-\infty}^{\infty}e^{-\frac{M^{2}}{2}}\textbf{\textrm{W}}''(P_{t}+(1+kM)P_{r})dM,\\
\frac{\partial^{2}}{\partial P_{r}^{2}}\bar{\textbf{\textrm{W}}} = \frac{1}{\sqrt{2\pi}}\int\limits_{-\infty}^{\infty}e^{-\frac{M^{2}}{2}}(1+kM)^{2}\textbf{\textrm{W}}''(P_{t}+(1+kM)P_{r})dM,\\
\frac{\partial^{2}}{\partial P_{r}\partial P_{t}}\bar{\textbf{\textrm{W}}} = \frac{1}{\sqrt{2\pi}}\int\limits_{-\infty}^{\infty}e^{-\frac{M^{2}}{2}}(1+kM)\textbf{\textrm{W}}''(P_{t}+(1+kM)P_{r})dM.
\end{array}
\end{equation}
\normalsize
Therefore based on the Cauchy–-Schwarz inequality and (\ref{partial deravitives of pr pt}), the determinant of the Hessian satisfies
\begin{displaymath}
\frac{\partial^{2}}{\partial P_{t}^{2}}\bar{\textbf{\textrm{W}}}\frac{\partial^{2}}{\partial P_{r}^{2}}\bar{\textbf{\textrm{W}}}-(\frac{\partial^{2}}{\partial P_{r}\partial P_{t}}\bar{\textbf{\textrm{W}}})^{2} > 0.
\end{displaymath}
In addition $\frac{\partial^{2}}{\partial P_{t}^{2}}\bar{\textbf{\textrm{W}}}>0$, so the Hessian is positive definite. $\bar{\textbf{\textrm{W}}}(P_{t},P_{r})$ is jointly convex of $P_{t},P_{r}$. \hfill\(\Box\)

\subsection{Double Markets Participation}\label{double market participation}
The single market operation is a straightforward one step optimization problem. In a more realistic model, the SBO participates in both the day-ahead and real time market to procure energy packets. As in Section \ref{single market participation}, the day ahead market allows the SBO to reserve a certain number of energy packets $P_{t}$ with purchasing unit cost of $k_{t}$. When the actual wind resource $P_{v}$ is revealed, the SBO can either sell back or purchase additional energy packets based on its need. If it sells back energy packets, it will be credited with $\gamma k_{t}$ per unit where $\gamma$ is the discount factor for diminished value of energy. If the SBO wishes to procure additional balancing energy, it will be purchased with higher unit price $k_{b}>k_{t}$. This is unknown in the day-ahead market, but the probability distribution $p(k_{b})$ can be modelled and is known to the SBO based on data history.

The double market participation is a two stage decision problem which can be solved backwards from the second to the first stage. In the second stage, namely the real time operation, the optimal strategy is to choose $x_{1}$ amount of packets from day-ahead reservation and $x_{2}$ from balancing energy to solve
\begin{equation}\label{real time solution}
\begin{aligned}
& \underset{x_{1},x_{2}}{\text{min}}
& & k_{b}x_{2} - \gamma k_{t}(P_{t}-x_{1})+\textbf{\textrm{W}}(x_{1}+x_{2}+P_{v}) \\
& \text{s.t.}
& & x_{1} \leq P_{t}, \hspace{3mm}x_{1},x_{2}\geq 0\\
\end{aligned}
\end{equation}
The second term is the credit the SBO will receive by selling $P_{t}-x_{1}$ energy packets back to the market. Denote the optimal value of (\ref{real time solution}), namely the real time optimal operation costs, as $\textbf{\textrm{R}}(P_{t},P_{v},k_{b})$, which is a function of day ahead purchase $P_{t}$, and the stochastic price $k_{b}$ and wind output $P_{v}$. The first stage, namely the day ahead optimal reservation for $P_{t}$ to hedge price and resource uncertainty, is to solve
\begin{equation}\label{day ahead planning}
\underset{P_{t}\geq 0}{\text{min}}\hspace{3mm} k_{t}P_{t}+\underset{P_{v},k_{b}}{E}[\textbf{\textrm{R}}(P_{t},P_{v},k_{b})].
\end{equation}
In (\ref{day ahead planning}) we assume that the SBO cannot optimize over $P_{r}$, this assumption will be relaxed in what follows. From standard sensitivity analysis, we note that for a fixed pair of $\{P_{v},k_{b}\}$, $\textbf{\textrm{R}}(P_{t},P_{v},k_{b})$ is a convex function of $P_{t}$. Therefore $\underset{P_{v},k_{b}}{E}[\textbf{\textrm{R}}(P_{t},P_{v},k_{b})]$, which is the expected value of the optimal real time operating cost overs price uncertainty $k_{b}$ and resource uncertainty $P_{v}$, is also convex. A first necessary order condition requires
\begin{displaymath}
k_{t}+\frac{d}{dP_{t}}\underset{P_{v},k_{b}}{E}[\textbf{\textrm{R}}(P_{t},P_{v},k_{b})]=0.
\end{displaymath}
Since the uncertainty is independent of $P_{t}$, we solve the equivalent problem
\begin{displaymath}
k_{t}+\underset{P_{v},k_{b}}{E}[\frac{d}{dP_{t}}\textbf{\textrm{R}}(P_{t},P_{v},k_{b})]=0.
\end{displaymath}
The Lagrangian function of the real time problem (\ref{real time solution}) is 
\begin{displaymath}
\begin{array}{ll}
L(x_{1},x_{2},\mu) = & k_{b}x_{2} - \gamma k_{t}(P_{t}-x_{1})+\textbf{\textrm{W}}(x_{1}+x_{2}+P_{v}) \\ & + \mu(x_{1} - P_{t}).
\end{array}
\end{displaymath}
Denoting $\mu^{\star}(P_{t},P_{v},k_{b})$ as the dual variable associated with the constraint $x_{1}\leq P_{t}$ at optimality, then
\begin{displaymath}
\frac{d}{dP_{t}}\textbf{\textrm{R}}(P_{t},P_{v},k_{t,h}) = -\gamma P_{t} - \mu^{\star}(P_{t},P_{v},k_{b}).
\end{displaymath}
Therefore the optimality condition is equivalent to
\begin{equation}\label{optimality for pt}
(1-\gamma)k_{t}-\underset{P_{v},k_{b}}{E}[\mu^{\star}(P_{t},P_{v},k_{b})]=0.
\end{equation}
Problem (\ref{optimality for pt}) can be solved iteratively by stochastic approximation with proper step size $\alpha(t)$ \cite{bottou1998online}. Proposition 7 formally establishes the almost sure convergence of $P_{t}$ to the optimum.

\textbf{Proposition 7} Denote $P_{t}^{\star}$ as the optimal solution to (\ref{day ahead planning}), and $\{P_{v}(i), k_{b}(i)\}$ as randomly generated sample uncertainty in the $i-$th iteration. If we update $P_{t}$ according to 
\begin{displaymath}
P_{t}(i+1)=P_{t}(i) - \alpha(i)[(1-\gamma)k_{t}-\mu^{\star}(P_{t}(i),P_{v}(i),k_{b}(i))]
\end{displaymath}
where $\alpha(i)=1/i$ and $\mu^{\star}(P_{t}(i),P_{v}(i),k_{b}(i))$ is the dual variable for constraint $x_{1}\leq P_{t}$ in (\ref{real time solution}) gotten at optimality, then $P_{t}(i)$ converges to $P_{t}^{\star}$ with probability 1 as $i\rightarrow\infty$.

\textit{Proof.} The proof is similar to Section 4.3--4.5 in \cite{bottou1998online}. \hfill\(\Box\)

A key assumption to guarantee the almost sure convergence is that the uncertainty in the second stage is independent of the decision in the first stage. The two stage problem becomes more difficult when the SBO can reserve and optimize over both $P_{t}$ and $P_{r}$. In this case, the uncertainty $P_{v}$ revealed in the second stage is a decision dependent variable based on $P_{r}$. For the day ahead decision, we solve
\begin{equation}\label{double market pt pr}
\begin{aligned}
& \underset{P_{t},P_{r}\geq 0}{\text{min}} 
& & k_{t}P_{t}+k_{r}P_{r}+\underset{P_{v},k_{b}}{E}[\textrm{\textbf{R}}(P_{t},P_{v},k_{b})]\\
& \text{s.t.}
& & P_{v}\sim N(P_{r},(kP_{r})^2)
\end{aligned}
\end{equation}
where $k_{r}$ is the unit cost for reserving wind energy, and $P_{v}$ is a Gaussian random variable depending on the decision $P_{r}$. The optimality condition for $P_{t}$ is the same as in (\ref{optimality for pt}). The optimality condition for $P_{r}$ can be derived by taking the derivative of the objective function in (\ref{double market pt pr})

\small
\begin{equation}\label{eq42}
\begin{array}{ll}
& k_{r}+\frac{\displaystyle\partial}{\displaystyle\partial P_{r}} \underset{P_{v},k_{b}}{E}\Big[\textbf{\textrm{R}}(P_{t},P_{v},k_{b})\Big]\\
=& k_{r} +  \underset{k_{b}}{E} \Big[\int\limits_{P_{v}} \textbf{\textrm{R}}(P_{t},P_{v},k_{b}) \frac{\displaystyle\partial}{\displaystyle\partial P_{r}} p(P_{v}|P_{r}) dP_{v}\Big]\\
=& k_{r} +  \underset{k_{b}}{E} \Big[\int\limits_{P_{v}} \textbf{\textrm{R}}(P_{t},P_{v},k_{b})
\frac{1}{P_{r}} (\frac{P_{v}(P_{v}-P_{r})}{(kP_{r})^{2}}-1) p(P_{v}|P_{r}) dP_{v}\Big] \\
= & k_{r} +  \underset{k_{b},P_{v}}{E} \Big[ \textbf{\textrm{R}}(P_{t},P_{v},k_{t,h}) \frac{1}{P_{r}} (\frac{P_{v}(P_{v}-P_{r})}{(kP_{r})^{2}}-1)\Big].
\end{array}
\end{equation}
\normalsize
The second equality is gotten since the conditional distribution is $p(P_{v}|P_{r})=\frac{1}{\sqrt{2\pi}kP_{r}}e^{-\frac{1}{2k^{2}}(P_{v}/P_{r}-1)^{2}}$, and  therefore 
\begin{displaymath}
\frac{\partial}{\partial P_{r}}p(P_{v}|P_{r}) = \frac{1}{P_{r}} (\frac{P_{v}(P_{v}-P_{r})}{(kP_{r})^{2}}-1) p(P_{v}|P_{r}).
\end{displaymath}
For notational simplicity, we let
\begin{equation}\label{f function}
f(P_{v},P_{r}) = \frac{1}{P_{r}} (\frac{P_{v}(P_{v}-P_{r})}{k^{2}P_{r}^{2}}-1)
\end{equation}
and substitute $f(P_{v},P_{r})$ into (\ref{eq42}) to get the optimality condition for $P_{r}$
\begin{equation}\label{optimality for pr}
k_{r}+\underset{P_{v},k_{b}}{E} \big[ \textbf{\textrm{R}}(P_{t},P_{v},k_{b}) f(P_{v},P_{r})\big]=0.
\end{equation}
Hence the day-ahead procurement can be gotten by jointly solving (\ref{optimality for pr}) and (\ref{optimality for pt}). Remember the almost sure convergence of the stochastic approximation algorithm requires that the uncertainty is independent of the control. This is satisfied for $P_{t}$ which will not affect uncertainties, but is partially satisfied for $P_{r}$ which will affect $P_{v}$. Therefore the stochastic update for $P_{t}$ and $P_{r}$ will not have the same structure, namely in updating $P_{r}$ we need to get rid off the affect of $P_{v}$ by taking multiple samples. 

We will present three stochastic approximation algorithms to solve for the optimal $P_{t}$ and $P_{r}$. The first algorithm shares the same algorithmic structure as Proposition 7 and tries to update $P_{t}$ and $P_{r}$ separately. For a fixed $P_{r}$, we update $P_{t}$ as in Proposition 7. For a fixed $P_{t}$, we update $P_{r}$ by evaluating the objective function's stochastic gradient with respect to $k_{b}$ uncertainty. This algorithm will guarantee to reach optimality almost surely, but the execution is slow due to time consuming evaluations in updating $P_{r}$ where we need to take multiple samples of $P_{v}$ to estimate the true gradient. The second algorithm, which simultaneously updates $P_{r}$ and $P_{t}$, is fast in reaching a neighbourhood of the optimal solution, but it will persist to oscillate and not converge. The third algorithm is a combination of the first two algorithms. It first uses the second algorithm to approach the neighbourhood of the optimal solution, and then it switches to the first algorithm to guarantee the almost sure convergence.

Before proceeding to the three algorithms, we first prove that the problem defined in (\ref{double market pt pr}) has the following property that guarantees the uniqueness of the optimal solution.

\textbf{Proposition 8} There is a unique local optimal $P_{t}$ ($P_{r}$) that minimizes (\ref{double market pt pr}), which is also global optimal for a fixed level of the other variable. 

\textit{Proof.} It is equivalent to prove that $\underset{P_{v},k_{b}}{E}[\textbf{\textrm{R}}(P_{t},P_{v},k_{b})]$ is a convex function of $P_{t}$ $(P_{r})$ while the other variable is fixed. Note that 
\begin{equation}\label{convex of pr pt}
\begin{array}{ll}
&\underset{P_{v},k_{b}}{E}\Big[\textbf{\textrm{R}}(P_{t},P_{v},k_{b})\Big]\\
=& \underset{k_{b}}{E}\Big[\int \textbf{\textrm{R}}(P_{t},P_{v},k_{b})p(P_{v}|P_{r}) dP_{v}\Big]\\
=& \underset{k_{b}}{E}\Big[\int \textbf{\textrm{R}}(P_{t},(1+kM)P_{r},k_{b})e^{-\frac{M^{2}}{2}} dM\Big]
\end{array}
\end{equation}
where the second equation is gotten by a change of variable $M=\frac{P_{v}-P_{r}}{kP_{r}}$. It is left to prove that $\textbf{\textrm{R}}(P_{t},(1+kM)P_{r},k_{b})$ is convex. In what follow, we prove the convexity of (\ref{convex of pr pt}) w.r.t $P_{r}$. The proof for $P_{t}$ follows similarly.

Note that the balancing energy price is strictly greater than day-ahead energy price, i.e. $k_{b}>k_{t}$. Therefore the SBO will have only three purchasing options $\textrm{(i)}$ purchase neither $x_{1},x_{2}$, $\textrm{(ii)}$ purchase $x_{2}=0$ and $x_{1}\leq P_{t}$, and $\textrm{(iii)}$ purchase $x_{1}=P_{t}$ and $x_{2}\geq 0$. The optimal real time operation cost will have the corresponding structure:
\begin{equation} \label{structure of real time cost}
\begin{array}{ll}
& \textbf{\textrm{R}}(P_{t},(1+kM)P_{r},k_{b}) \\
= & \Bigg\{
\begin{array}{l}
-\gamma k_{t}P_{t} + \textbf{\textrm{W}}((1+kM)P_{r}), \textrm{if (i)} \\
-\gamma k_{t}(P_{t}-x_{1}) + \textbf{\textrm{W}}(x_{1}+(1+kM)P_{r}), \textrm{if (ii)}\\
k_{b}x_{2} + \textbf{\textrm{W}}(P_{t}+x_{2}+(1+kM)P_{r}), \textrm{if (iii)}
\end{array}
\end{array}
\end{equation}
Namely the real time cost is a piece-wise continuous function having three parts corresponding to (\ref{structure of real time cost}). Since $\textbf{\textrm{W}}(\cdot)$ is convex, each piece-wise function is convex. In addition, both sub-gradients exist at the points that joint cases $\textrm{(i)},\textrm{(ii)}$ and cases $\textrm{(ii)},\textrm{(iii)}$. Therefore the piecewise function is convex in $P_{r}$. \hfill\(\Box\)

Algorithm 1 below is based on Proposition 8 that indicates that there is a unique optimal solution to update either $P_{t}$ or $P_{r}$ while having the other fixed. According to Proposition 7, if the uncertainty is independent of the decision variable, stochastic approximation can be used to find the optimal solution. This indicates that we can iteratively update $P_{t}$ and $P_{r}$ to approach the global optimal solution.

\underline{\textbf{Algorithm 1}}

(1) Set initial value of $P_{t}(0)$ and $P_{r}(0)$. Set $j=0$.

(2) {\bf Update $\mathbf{P_{t}}$ for fixed $\mathbf{P_{r}}$}. Let $P_{t}^{0}=P_{t}(j)$ and for $i=0,1,...,M$ with some large $M$ generate i.i.d samples of $k_{b}^{i}$, and $P_{v}^{i}$ based on $P_{r}(j)$. Solve (\ref{real time solution}) to get the dual variable $\mu^{\star}(P_{t}^{i},P_{v}^{i},k_{b}^{i})$. Update $P_{t}^{i}$ according to
\begin{displaymath}
P^{i+1}_{t} = P^{i}_{t} - \alpha_{i}((1-\gamma)k_{t} - \mu^{\star}(P_{t}^{i},P_{v}^{i},k_{b}^{i})).
\end{displaymath} 
Let $P_{t}(j+1)=P_{t}^{M}$ after the $M-$th iteration.

(3) {\bf Update $\mathbf{P_{r}}$ for fixed $\mathbf{P_{t}}$}. Let $P_{r}^{0}=P_{r}(j)$ and for $i=0,1,...,M$ with some large $M$ generate i.i.d samples of $k_{b}^{i}$. Solve (\ref{real time solution}) under different $P_{v}$ to get the dual variable $\textrm{\textbf{R}}(P_{t}(j+1),P_{v},k_{b}^{i})$. Calculate the derivative w.r.t $P_{r}$ by the following integration
\begin{equation}\label{integrate to get derivative of pr}
\int\limits_{P_{v}} \textbf{\textrm{R}}(P_{t},P_{v},k_{b}) f(P_{v},P_{r}^{i}) p(P_{v}|P_{r}^{i}) dP_{v},
\end{equation}
where $f(\cdot,\cdot)$ is defined in (\ref{f function}), and update $P_{r}^{i}$ according to
\begin{displaymath}
P^{i+1}_{r} = P^{i}_{r} - \alpha_{i}\Big[k_{r} + \int\limits_{P_{v}} \textbf{\textrm{R}}(P_{t},P_{v},k_{b}) f(P_{v},P_{r}^{i}) p(P_{v}|P_{r}^{i}) dP_{v}\Big].
\end{displaymath} 
Let $P_{r}(j+1)=P_{r}^{M}$ after the $M-$th iteration.

\begin{figure}[htb!]
\centering
\subfigure[Algorithm 1 favors almost sure convergence.]{\includegraphics[width = 7cm]{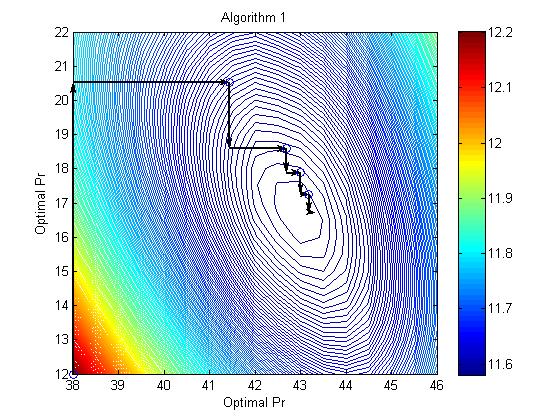}\label{fig:SubLabel1}}
\subfigure[Algorithm 2 favors computational efficiency.]{\includegraphics[width = 7cm]{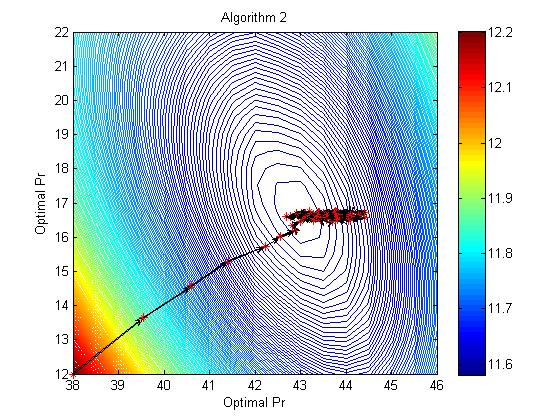}\label{fig:SubLabel2}}
\subfigure[Algorithm 3 combines the advantages.]{\includegraphics[width = 7cm]{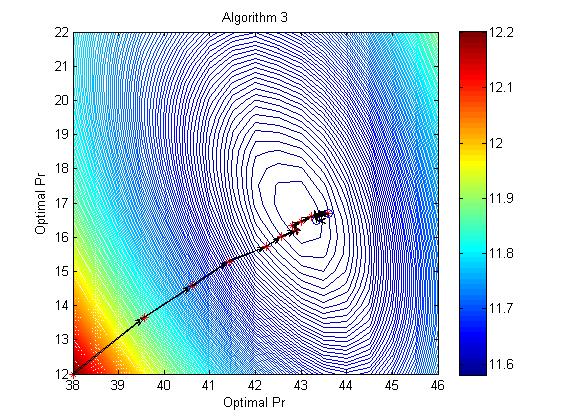}\label{fig:SubLabel3}}
\caption{ (a) Almost sure convergence of $P_{t}$ and $P_{r}$ generated by stochastic approximation Algorithm 1. In each iteration two sub-problems are solved. We first fix $P_{r}$ to get the optimal $P_{t}$, and then fix $P_{t}$ to get the optimal $P_{r}$. Almost sure convergence to the optimal is guaranteed for both sub-problems. In the end global optimality is achieved by iteratively solving the two sub-problems. (b) Algorithm 2 updates both $P_{t}$ and $P_{r}$ at the same iteration based on random samples of $k_{b}$ and $P_{v}$. In this algorithm the generated sequence of $P_{t}^{i}$ will converge to the near optimal solution $P_{t}^{\star}$, but $P_{r}^{i}$ will persist in oscillating around the optimal solution $P_{r}^{\star}$. (c) Algorithm 3 first adopts Algorithm 2 to update both $P_{t}$ and $P_{r}$ at the same iteration based on random samples of $k_{b}$ and $P_{v}$ until $P_{t}$ converges to $P_{t}(0)$ with some $P_{r}(0)$. Using $P_{t}(0)$ and $P_{r}(0)$ as a starting point, it then switches to Algorithm 1 to separately update $P_{r}(i)$ and $P_{t}(i)$ one at a time. Within few iterations (typically less than 3) of separate updating, both $P_{t}(i)$ and $P_{r}(i)$ converge.}
\label{fig:MainLabel}
\end{figure}

(4) {\bf Convergence Criterion}. If $|P_{t}(j)-P_{t}(j+1)|<\epsilon$ and $|P_{r}(j)-P_{r}(j+1)|<\epsilon$, return the almost sure optimal solution $P_{t}^{\star}=P_{t}(j+1)$ and $P_{r}^{\star}=P_{r}(j+1)$. Otherwise increase $j$ by 1 and go to step (2).

Fig. \ref{fig:SubLabel1} shows the trajectory generated by Algorithm 1. With sufficiently large $M$ chosen in step (2) and (3), this algorithm will find the optimal solution for $P_{t}$ and $P_{r}$. Algorithm 1 will converge almost surely to the global optimal $P_{t}^{\star}$ and $P_{r}^{\star}$ since step (2) and step (3) both converge to the almost surely optimal solution at the corresponding steps and that the problem itself is jointly convex of $\{P_{t},P_{r}\}$. 

One issue for Algorithm 1 is the computational effort needed in step (3) to accurately calculate the derivative of $P_{r}$. We need to consider all possible realizations of $P_{v}$, solve (\ref{real time solution}) for each $P_{v}$, and integrate (\ref{integrate to get derivative of pr}). The second algorithm avoids the explicit calculation of (\ref{integrate to get derivative of pr}). Instead it updates $P_{t}$ and $P_{r}$ at the same iteration based on samples of $k_{b}$ and $P_{v}$. 

\underline{\textbf{Algorithm 2}}

(1) Set initial value of $P_{t}^{0}$ and $P_{r}^{0}$. 

(2) {\bf Update $\mathbf{P_{r}}$ and $\mathbf{P_{t}}$}. For $i=0,1,...,M$, generate $k_{b}^{i}$ and $P_{v}^{i}$ based on $P_{r}^{i}$. Solve (\ref{real time solution}) to get $\textbf{\textrm{R}}(P_{t}^{i},P_{v}^{i},k_{b}^{i})$ and $\mu^{\star}(P_{t}^{i},P_{v}^{i},k_{b}^{i})$. Update $P_{t}^{i}$ and $P_{r}^{i}$
\begin{displaymath}
\begin{array}{c}
P_{t}^{i+1} = P_{t}^{i} - \alpha_{i}\Big[ (1-\gamma)k_{t}-\mu^{\star}(P_{t}^{i},P_{v}^{i},k_{b}^{i}) \Big]\\
P_{r}^{i+1} = P_{r}^{i} - \alpha_{i}\Big[ k_{r}+ \textbf{\textrm{R}}(P_{t}^{i},P_{v}^{i},k_{b}^{i})f(P_{v}^{i},P_{r}^{i}) \Big]
\end{array}.
\end{displaymath}
Return $P_{t}^{\star}=P_{t}^{M}$ and $P_{r}^{\star}=P_{r}^{M}$ as the near optimal solution.

Fig. \ref{fig:SubLabel2} shows the trajectory generated by Algorithm 2 where the solution converges to the optimal in $P_{t}$ direction, but will persist to oscillate inside a set containing the optimal solution $P_{r}^{\star}$. Algorithm 2 is fast compared with Algorithm 1 since each time we solve (\ref{real time solution}), we get the optimal cost as well as the dual variable that are used to update both $P_{t}$ and $P_{r}$. However, after a certain number of iterations this algorithm becomes inefficient since $P_{r}$ will oscillate while $P_{t}$ will stay at the same level.

The above two algorithms show a trade off between the almost sure convergence and the speed (computational effort) in calculating the sample derivative of $P_{r}$. Since the second algorithm can successfully drive the trajectory to a region that is close to almost sure optimality, we can combine the advantages of the two algorithms such that we first speed to the neighbourhood of optimality and then switch to Algorithm 1 to avoid oscillation of $P_{r}$. This inspires the Algorithm 3 which is given as follows.

\underline{\textbf{Algorithm 3}}

(1) Run Algorithm 2 until $P_{t}$ converges. Typically the converged value is close to the optimal solution $P_{t}^{\star}$. Denote the solution by $P_{t}(0)$ and $P_{r}(0)$. Set $j=0$.

(2) Run step (3) in Algorithm 1 to update $P_{r}(j)$ to $P_{r}(j+1)$ while fixing $P_{t}(j)$.

(3) Run step (2) in Algorithm 1 to update $P_{t}(j)$ to $P_{t}(j+1)$ while fixing $P_{r}(j+1)$.

(4) Check convergence of $P_{r}(j+1)$ and $P_{t}(j+1)$ according to convergence criterion (4). If not, increase $j$ by 1 go back to step (2).

Fig. \ref{fig:SubLabel3} shows the trajectory generated by Algorithm 3. The initial trajectory is similarly generated as in Algorithm 2. When $P_{t}$ converges, we switch to Algorithm 1 to avoid the oscillation phenomenon observed in Algorithm 2. Since $P_{t}(0)$ gotten at the end of step (1) is very close to the real optimal solution $P_{t}^{\star}$, we can expect a few iterations between step (2) to step (4) before the final convergence.

\section{Control and Market Aspects for the Binary Information PDLC}\label{Control and Market Aspects for the Binary Information PDLC}
We would like address properties of the binary information based PDLC framework from both control and market perspectives. In terms of control, we seek to understand the fundamental trade off between the benefit of providing flexible energy provision to consumers and the costs of facing larger system instability. In terms of markets, we will address the effect of wind energy on market contracts.

\subsection{Demand Uncertainty vs. Consumer's Utility}
It is well known that a trade off exists between providing consumers with higher flexibility and ensuring a better controlled and predictable system \cite{5643088}. For example, in numerous papers price based indirect load control frameworks have been proposed that allow consumers to independently decide their energy preferences based on real time price signals. However, few demand response programs nowadays actually exercise such practice due to the high unpredictability of the system response. DLC mechanisms, on the other hand, restrict the allowable actions by consumers to regulate their appliances in order to achieve accurate and robust performance of the distribution network. In our binary information based PDLC framework, there are two parameters governing the trade off between flexibility and controllability -- the number of reserved energy packets $m$ and packet length (decision duration) $\delta$. Higher level of $m$ or smaller value of $\delta$ provides flexibility as the system is enabled with a higher maximum capacity and more frequent decision making possibility 

\begin{figure}[htb]
\centering
\subfigure[Variance of demand reflects the controllability of the PDLC.]{\includegraphics[width = 7cm]{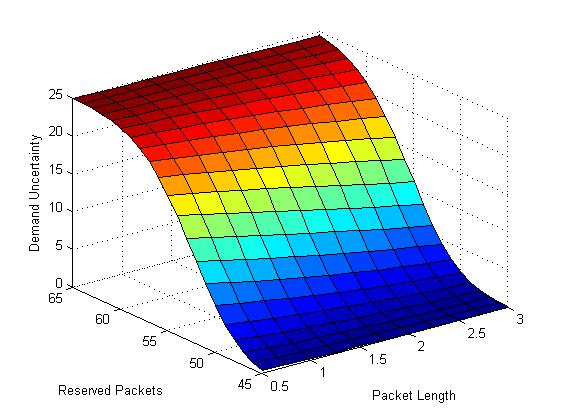}\label{fig:Variance}}
\subfigure[Mean Waiting Time reflects consumers' dis-comfort of having temperature rise beyond the desired range.]{\includegraphics[width = 7cm]{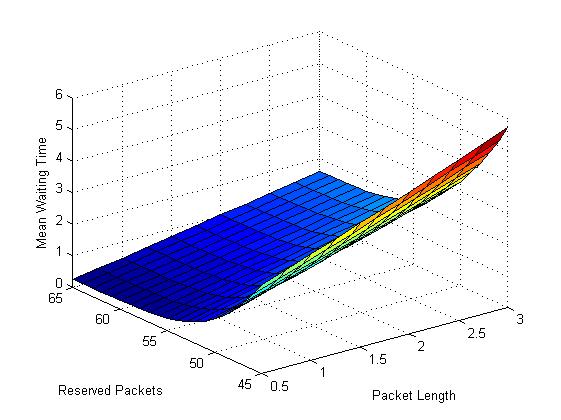}\label{fig:MWT}}
\caption{ Trade off between flexibility and controllability. System uncertainty would monotonically increase as either we reduce decision making period, i.e. reducing packet length $\delta$, or we increase the allowable maximum system capacity, i.e. increasing amount of packet reservation $m$. Consumers flexibility will have opposite characteristics to the system uncertainty.}
\label{fig:MainLabel2}
\end{figure}

We numerically illustrate this trade off in Fig. \ref{fig:MainLabel2} for system uncertainty $\textrm{Var}(n(m,\delta))$ and mean waiting time $W(m,\delta)$. It can be seen from Fig. \ref{fig:Variance} that the variance of the packet consumption is affected by (i) the maximum reservation amount of energy packets $m$ and (ii) the duration of packet length $\delta$. The former has a greater effect. Regardless of $\delta$, demand uncertainty is static when we either provide absolute freedom (with large $m$) or control (with small $m$) to consumers. We also observe that $\textrm{Var}(n(m,\delta))$ increases linearly with respect to values of $m$ around the optimal procurement level.

The number of reserved packets $m$ also has a greater effects on $W(m,\delta)$ as shown in Fig. (\ref{fig:MWT}). With a moderate amount of reservation, $W(m,\delta)$ can be reduced to a value less than 20 seconds. Considering that the duty cycle on and off time are typically around 10 minutes, delaying 20 seconds will only increase the room temperature beyond the boundary by 5\% of the comfort band width. If additional reduction of $W(m,\delta)$ is desired, the SBO can choose to have smaller $\delta$ which will have larger effect than reserving additional packets. The choice of small $\delta$ does have small effect of increasing demand uncertainty.

\subsection{Wind Energy Impacts}
We would like to quantify the value of wind energy in market operations. The optimal usage of wind energy is to balance between: (i) the benefit it brings to the system welfare because of its low operating costs, and (ii) the uncertainty it brings to the real time operation such that more operating reservations, both traditional and renewable, are needed. One important aspect is to find how wind quality will affect the optimal procurement of wind energy, and its consequential effects on potential contacts between wind producers and the SBO. Higher uncertainty in wind availability will discourage the demand side from using wind resources in favor of continued reliance on traditional energy sources. In addition, the operation costs of wind $k_{r}$ will also have an impact on the energy contract between the SBO and the wind farm.

Table \ref{tab:1} and Table \ref{tab:2} show the optimal energy procurement for $P_{r}$ and $P_{t}$ as a function of wind quality and cost with coefficient of variations $k$ and wind operational costs $k_{r}$. We find a linear dependence between $k$ and the contracted wind output $P_{r}$ where $P_{r}$ decreases as $k$ increases. $k_{r}$ also has a minor effect on the optimal $P_{r}$ as long as $k_{r}$ is relatively lower than the costs of day-ahead energy reservation. As expected, the demand for $P_{t}$ will increase in a way that is complementary to a decreased demand for $P_{r}$. It is clear that total amount of energy reservation, i.e. $P_{t}+P_{r}$ will increase as either the wind quality deteriorates or wind costs decrease. 

\begin{table}
\centering
\caption{Optimal Contract of the Wind Output as Wind Price or Quality Changes}\label{tab:1} 
\begin{tabular}{|c|c|c|c|c|c|}
\hline $k \backslash k_{r}$ & 0.02 & 0.04 & 0.06 & 0.08 & 0.10 \\ 
\hline 0.05 & 52.66 & 52.37 & 51.97 & 51.52 & 50.77 \\ 
\hline 0.10 & 49.88 & 49.43 & 48.86 & 48.11 & 47.52 \\ 
\hline 0.15 & 47.05 & 46.52 & 45.87 & 44.91 & 43.93 \\ 
\hline 0.20 & 44.41 & 43.68 & 42.92 & 41.97 & 40.84 \\ 
\hline 0.25 & 41.75 & 41.03 & 40.20 & 39.17 & 38.14 \\ 
\hline 0.30 & 39.52 & 38.61 & 37.80 & 36.86 & 35.59 \\ 
\hline 
\end{tabular} 
\end{table}

\begin{table}
\centering
\caption{Optimal Contract of Traditional Energy as Wind Price or Quality Changes}\label{tab:2} 
\begin{tabular}{|c|c|c|c|c|c|}
\hline $k \backslash k_{r}$ & 0.02 & 0.04 & 0.06 & 0.08 & 0.10 \\ 
\hline 0.05 & 3.50 & 3.72 & 4.14 & 4.53 & 4.95 \\ 
\hline 0.10 & 7.89 & 8.08 & 8.86 & 9.56 & 9.84 \\ 
\hline 0.15 & 12.14 & 12.31 & 12.84 & 13.91 & 14.77 \\ 
\hline 0.20 & 15.95 & 16.49 & 17.09 & 17.89 & 18.93 \\ 
\hline 0.25 & 20.11 & 20.47 & 20.90 & 21.78 & 21.66 \\ 
\hline 0.30 & 22.48 & 23.61 & 23.48 & 24.51 & 25.65 \\ 
\hline 
\end{tabular} 
\end{table}

\section{Conclusion and Future Path}\label{conclusion}
Smart buildings account for the greatest share of microgrid energy demand. They also provide the greatest opportunity for management and control. A new operating protocol, called \textit{packetized direct load control} (PDLC), has been introduced and proposed for various communication and control settings for the smart building operation. Two levels of possible communications are considered that comprise (i) an ideal scenario where consumers' allow the operator to access their real time full information with an erasure free channel, and (ii) a constrained information exchange scenario where the operator has limited access regarding the binary desirability of consumers electricity preference. We show the fundamental trade off between achieving controllability of the system and endowing flexibility to consumers within the PDLC as the operator varies the reservation capacity or decision interval duration. This trade off is further mathematically defined and unified into either an energy metric or a consumer welfare metric for optimization purposes. 

Based on these metrics, the concept of PDLC is embedded into market settings where we consider the SBO's participation in the cascading of day-ahead and real time markets for optimal energy procurement. We have proposed three algorithms that solve the corresponding stochastic approximation problem and whose result gives the optimal energy management solution to purchase a mixed portfolio of traditional and renewable (wind) energy. Along the course, we proved the uniqueness of the optimal solution and the almost sure convergence of our algorithms. The impact of wind energy with respect to its quality and costs is numerically addressed.

We believe that the PDLC framework can be fit into broader settings in optimal control of demand response and optimal operation of electricity markets. This paper serves as a seminal work discussing the role of PDLC operation in the energy market. One direct extension is to consider the PDLC in reserve markets where the operator can flexibly modulate aggregate consumption up and down by rationing energy packets based on the needs of consumers. The role of information is a critical enabler of optimized performance of building energy systems.

A further extension along the path is to explore the possibility of using \textit{energy packets} to reach higher operational granularity that is not necessarily restricted within the direct load control framework. Price based protocols would easily be adapted where consumers can potentially sign packet-based contracts containing various choices of packet duration depending on their beliefs on prices and their need for energy. Similar to the above, packet contract duration and allowable reservation capacity will play a role in balancing system controllability and consumer flexibility.


%

%

\ifCLASSOPTIONcaptionsoff
  \newpage
\fi



%
\bibliography{references}{}
\bibliographystyle{unsrt}
%

%





\end{document}